\documentclass{mem}
\usepackage{natbib}\usepackage{txfonts}\usepackage{balance}
\usepackage{graphicx}
\usepackage[a4paper]{hyperref}
\idline{?}{1}
\begin{document}
\def\teff{$T\rm_{eff }$}
\def\kms{$\mathrm {km s}^{-1}$}

\title{Progress in understanding and exploiting stellar oscillation
spectra }

   \subtitle{}

\author{
W.A.\, Dziembowski
          }

  \offprints{W.A. Dziembowski}

\institute{Warsaw University Observatory, Al. Ujazdowskie 4,
00-478 Warsaw, Poland \and Copernicus Astronomical Center, ul.
Bartycka 18, 00-716 Warsaw, Poland }

\authorrunning{ Dziembowski }

\titlerunning{Stellar oscillation spectra }

\abstract{

Rich oscillation spectra of dwarf-like pulsators
contain a wealth of information about the object interiors and, in
particular, about macroscopic transport processes, which is the
most difficult aspect of stellar physics.
Examples of extracting such information
from data on solar-like and  opacity driven pulsators are given.
Problems in understanding new oscillation spectra are discussed.
Importance of employing various data on excited mode is
emphasized.

 \keywords{Stars: oscillations -- Stars: interiors --
Stars: individual $\alpha$ Cen, $\eta$ Boo, PG 1605+072, FG Vir, $\nu$ Eri} } \maketitle{}

\section{Introduction}

Multimode low-amplitude variability is the most common form of
stellar pulsation. Though  typical for dwarfs, it is found also in
giants. Much hope is attached to observations of objects
exhibiting this form of pulsation because each frequency of a
known mode is an independent high precision probe of stellar
interior. However, mode identification in oscillation spectra is
often a difficult task, which requires additional observables,
such as amplitudes in various passbands, or theoretical arguments
based on stellar model and oscillation calculations. Thus, the
impressive progress in detecting low amplitude variability and
measuring individual mode frequencies was not yet reciprocated in
wider exploitation of the new data.

The most important application of frequency data is deriving
constraints on stellar interior physics and, more specifically on
slow macroscopic flows, which are so important for chemical
evolution of stars, but so difficult for laboratory or numerical
simulations. Successes of helioseismology in probing flows are
encouraging but we have to keep in mind the difference in number
of measured frequencies and the role of high-degree p-modes, which
will not soon be detected in distant stars. The method of
asteroseismic probing is the construction of {\it seismic models},
that is, models whose oscillation frequencies reproduce data and
whose global parameters are consistent with non-seismic data. Free
parameters describing internal structure and theoretical
uncertainties in stellar modeling are treated as adjustable
quantities. Typically, there is also a freedom in oscillation mode
identification.

Multimode stellar pulsation is a complicated phenomenon, only
partially understood. The current status of our understanding is
briefly reviewed in the next section. The rest of this paper is
focused on data interpretation for five objects. I believe that
the problems encountered in these cases are sufficiently diverse
to show the main research direction where progress has been
achieved, and where we still  have outstanding questions to
answer.

\section{Physics of multimode pulsation}

Multimode pulsation may arise in two distinct ways. One is a
stochastic excitation by turbulent convection in outer layers.
Modes driven in this way are termed {\it solar-like oscillations}.
The other, which may be referred to as {\it self-excited
oscillations} arises as a result of linear instability of
oscillatory modes. We usually know which of the two driving
mechanisms operates in the star. Only in the case of the common
low amplitude variability in red giants, the issue may still be
regarded as open. For most of self-excited mode pulsators, we
believe that we understand the physics of driving effect, which
most often is the usual opacity mechanism working in different
opacity bump zones. However, our understanding of observed
oscillation spectra is still far from being satisfactory, as it
requires nonlinear theory implying much higher level of
difficulties in modeling.

The goal of reproducing the gross features, such as frequency
dependence of peak amplitudes and widths of the observed
oscillation spectra for solar-like pulsators is not achieved yet.
This has been clearly seen in the attempt by \citet{sam} to apply
the stochastic excitation theory scaled on the sun to $\alpha$
Centauri. Observed power of individual modes turned out larger
than predicted  by large factors (2 to 7). Furthermore, there is a
significant discrepancy between observed and predicted widths of
the peaks  \citep{bed}.

Very little is known about nonlinear development of unstable
modes. We can identify possible amplitude-limiting effects, such
as changes in the mean structure induced by pulsation or resonant
coupling to stable modes, but model calculations taking to account
both effects have yet to be done. Therefore, we still cannot
explain why, in spite of essentially the same driving mechanism,
certain stars choose high-amplitude monomode pulsation while
others choose low-amplitude multimode pulsation. Observations tell
us that the difference is related to evolutionary status. The
theory suggests the clue may be very different properties of
nonradial modes which change drastically with the advance of
evolution. However, realistic modeling of mode interaction is
still ahead of us.

\section{Solar-like pulsators}

Oscillation spectrum of $\alpha$ Centauri A, obtained by
\citet{bed}, resembles first spectra of solar low-degree
modes obtained about 25 years ago.
Like in the case of the solar spectra,
mode identification in the $\alpha$ Cen spectrum is unambiguous.
There are also oscillation data on the lower mass component of
the binary $\alpha$ Cen system. In addition, we have also excellent constraints on
the two star masses and radii from orbital and interferometric
data.

In a recent work, \citet{mm} present a comprehensive survey of
models subject to seismic and non-seismic constraints. The
conclusions are somewhat disappointing for asteroseismology. A
number of models were found, which within the observational errors
reproduce all the data. There is a marginally significant
discrepancy in the small separations $d_{01}$, which are sensitive
to structure of the interior. The most interesting result, namely
the determination of the convection parameter, $\alpha_{\rm con}$,
for two stars, relies entirely on the non-seismic constraints.

This is not the end of $\alpha$ Cen seismology. More accurate
frequency measurements should yield precise values of $d_{01}$. If
the discrepancy is confirmed, we will have an interesting problem
to solve. The problem we have now is the conflict between expected
and observed mode amplitudes and life-times mentioned in the
previous section.

More promising but much harder to interpret are spectra for more
massive and more evolved objects, Procyon and $\eta$ Bootis. In
these cases we have a prospect for precise probing of the deep
interior employing mixed mode frequencies. With progress of
stellar evolution, the g-mode frequency increase leads to the {\it
avoided crossing} phenomenon, which in the context of modeling
$\eta$ Bootis was discussed by \citet{dima}. Occurrence of the
mixed mode destroys the rhythm of p-modes which complicates
interpretation of the spectrum but the reward is the high
sensitivity of the avoiding crossing frequencies to the structure
of the deep interior.

Prospect for sounding interior of intermediate mass stars
($1.2-1.8 M_\odot$) is appealing because, unlike in more massive
objects, the convective core recedes during  main sequence
evolution. This may lead to the formation of a semiconvective
zone, or a nearly discontinuous rise of hydrogen abundance
(\citet{pop} and reference herein). It is also possible that the
distribution of elements is smoothed by overshooting or by the
rotation induced mixing. We do not know which is true.

$\eta$ Boo is the first distant star where solar-like oscillation
were definitely detected and mode frequencies were measured
\citep{kje}. Already in this first data there was some evidence
for mode departing from ordinary p-mode pattern. There are newer
oscillation spectra for this star but still there is no credible
measurement of the avoided crossing frequency.

\section{Seismic models without observational constraints on modes}

In recent years there has been considerable interest in
observations and interpretation of oscillations in pulsating
B-type subdwarfs. The objects, which are believed to belong to the extreme horizontal
branch, are indeed attractive targets for asteroseismology.
They are truly multimodal and there is a prospect for detection of traces of the
previous evolutionary phases in the internal structure.

In Fig. \ref{wdf1}, we may see a schematic oscillation spectrum
for the sdB pulsator PG 1605+072 from \citet{charpinet}. The
spectrum lacks repeating spacings, which could be used as a clue
to mode identification. The authors based their mode
identification on the double-optimization method developed by the
Montreal group and used in their earlier works cited in the paper.
The method consists in considering oscillation frequencies in a
four parameter family of stellar envelopes in thermal and
diffusion equilibrium. The parameters are the total mass, $M$, the
mass of the hydrogen-rich outer layer, $M_H$, the effective
temperature, $T_{\rm eff}$, and the surface gravity, $g$. For each
model a combination of low-degree ($\ell\le3$) modes is found
which minimizes the weighted distance between the observed and
calculated frequencies, $$\chi^2=\sum_k\left({\nu_{k,{\rm
obs}}-\nu_{k,{\rm cal}}\over\sigma_k}\right)^2.$$ The subsequent
minimization of $\chi^2$ with respect of the stellar parameters
leads to identification of modes and best values of $M$, $M_H$,
$T_{\rm eff}$, and $g$. The values of the last two parameters are
known from spectroscopy  which, in the case of PG 1605+072,
allowed to choose between two $\chi^2$ minima of a comparable
depth. The parameters which may be determined only by means of
seismology are $M_H$ and (in this case) $M$. Only with these two
parameters determined we have sufficient constraints on the
evolutionary history of the objects.

Except of the upper limit of $\ell$, which comes from the
visibility argument, the set of considered modes was constrained
by the condition of mode instability, which sets an upper limit on
mode order, $n$. An argument was given that rotation is negligible
hence the azimuthal order, $m$, was irrelevant. However, the
argument for slow rotation, which is the lack of side peaks, is
not satisfactory because at the level of linear theory modes of
different $m$ are independent and not all multiplet components
must be excited.

All the peaks in the spectrum are explained in terms of unstable low order p-modes.
The instability range extends to much higher orders ($n=6$ at $\ell=0$).
The fact that all detected modes may be associated with unstable modes
is very encouraging and must be regarded as a support for
models assuming chemical element distribution determined by the
diffusion equilibrium with radiative levitation taken into
account. The fact that not all unstable low-degree modes are
detected is typical for stars having a large number of unstable
modes. However, some aspects of the proposed mode
identification are causing concern.

It is difficult to justify disregarding modes with $\ell>3$ in the
situation when three of the nine peaks were identified as $\ell=3$
and one of them is among the dominant ones. The large rise of the
cancellation effect due to disc averaging occurs between $\ell=2$
and 3. Between 3 and 4, the rise is much smaller and, in certain
cases, the trend may be even reversed. Therefore, it is  very
important to get an independent assessment of the mode degrees.

\begin{figure}[]
\resizebox{\hsize}{!}{\includegraphics[clip=true]{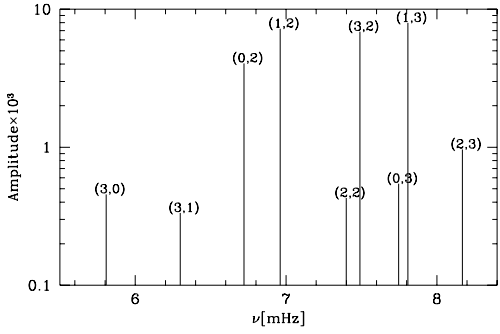}}
\caption{ \footnotesize Oscillation spectrum of PG 1605+072. The
numbers in the bracket give  $(\ell, n)$ - the  degree and radial
order of the mode. The data and mode identifications are from
\citet{charpinet}. At $\ell=0$ the mode count begins here with
$n=1$.\label{pg}}
\end{figure}

\section{Application of amplitude and phase data}

The oscillation spectrum of the $\delta$ Scuti star FG Virginis,
containing 67 independent frequencies obtained by \citet{breger},
presents the greatest challenge to theory. In spite of the large
number of peaks, the spectrum lacks repeating spacings. However,
unlike for sdB pulsators, we have accurate amplitude and phases
for a number of peaks. Such data allow, at least in principle, to
determine $\ell$ and $m$ values of the modes associated with the
peaks. Inference on the $\ell$ value is fairly straightforward and
may be done in various ways. Here I briefly outline the way
applied recently to the FG Vir data by \citet{jdda} which in
addition to $\ell$ yields additional constraints on the star. As
in all other methods, it is assumed that the atmosphere remains
plane-parallel and in thermal equilibrium during pulsation and
that the changes in the atmospheric parameters, $\delta{\cal
F}_{\rm bol}$ and $\delta g$, are related to the displacement of
the photospere, which is given in the form $$\delta
r(R,\theta,\varphi)= R {\rm Re}\{ \varepsilon Y_\ell^m {\rm
e}^{-{\rm i}\omega t}\}, $$ where $\varepsilon$ is a complex
unknown. Then, we have $$\delta {\cal F}_{\rm bol}={\cal F}_{\rm
bol} {\rm Re}(\varepsilon f Y_\ell^m {\rm e} ^{-{\rm i} \omega
t})$$ and $$\delta g = - \left( 2 + \frac{\omega^2 R^3}{G M}
\right) \frac{\delta r}{R},$$ where $f$, another complex unknown,
is a new seismic observable of specific diagnostic properties.

Light amplitude and phases in individual passbands are calculated
with the atmosphere model data calculated around specified ${\cal
F}_{\rm bol}$ and $g$. Amplitude and phase of the disc-averaged
radial velocity variation are determined by $\delta r$ and the
limb-darkening law. The problem is cast in the form of the linear
relations between measured complex amplitudes and the unknowns
${\tilde\varepsilon}$ and ${\tilde\varepsilon}f$, where
${\tilde\varepsilon} = \varepsilon Y^m_{\ell}(i,0)$.  The
coefficients in this relations depend on the central values of the
atmospheric parameters and on $\ell$ but not on $m$ nor on the
aspect angle, $i$. The values of $\ell$, ${\tilde\varepsilon}$ and
${\tilde\varepsilon}f$ are determined by $\chi^2(\ell)$
minimization.

For all the 12 modes with accurate photometric and spectroscopic
data, the degrees $\ell>2$ could be be rejected at a high
confidence level. In half of the cases the identification was
unique at the 80\% confidence level. In two cases the most likely
values are $\ell=0$. In addition, fitting the dominant peak sets
stringent limits on the allowed $T_{\rm eff}$ range. Future
seismic models of FG Vir should take into account all these
constraints in a probabilistic form.

At the present stage, the only direct seismic probe of the star
interior came from the $f$'s determined for the twelve modes.
These quantities probe only the outer layers including the
hydrogen and helium ionization zones, where convection carries
part of the energy flux. Thus, a comparison of the values derived
from data with those determined from  linear nonadiabatic
calculations of stellar oscillation yields a constraint on
modeling convection. \citet{jdda} showed that models assuming
inefficient convection ($\alpha_{\rm conv}\le0.5$) reproduce very
well observational values across the the whole frequency range
extending from 9.2 to 24.2 c/d covered the twelve modes.

We understand certain aspects of the FG Vir oscillation spectrum.
The dominant peaks are explained in terms of low-degree modes,
which all are found unstable in the models consistent with the
data. However, there are other significant peaks of which majority
must have $\ell>2$ and most likely $\ell>>2$
(Daszy\'nska-Daszkiewicz et al., in this volume). Model
calculations predict p-mode and low-$\ell$ g-mode instability of
frequencies from about 5 c/d to about 30 c/d. The whole range,
with varying density, is populated by the observed peaks. There is
also a number of peaks with amplitudes $\sim 0.2$ mmag detected
between 30 and 45 c/d which could not be interpret as second order
peaks of lower frequency modes. Only f-modes with $\ell>70$ are
unstable in this high frequency range. Without nonlinear
calculations we cannot tell whether this excitation of such modes
is a plausible interpretation. Another puzzling aspect of the
spectrum pointed out by Breger \& Pamyatnykh (in this volume) is
the occurrence of very close peaks near frequencies expected for
consecutive radial modes.

\section{Seismology of $\beta$ Cephei stars}

For years, the main interest in $\beta$ Cep pulsation focused on
the search of driving mechanism. After the new opacity data
essentially have solved the problem, the interest moved toward
seismic sounding of the stars. New extensive multisite
observations of several beta Cephei, revealed much richer
oscillation spectra than has been known before. The new data
enabled deriving reliable constraints on star interior structure
and rotation but also created some problems that await solution.

\begin{figure}[t!]
\resizebox{\hsize}{!}{\includegraphics[clip=true]{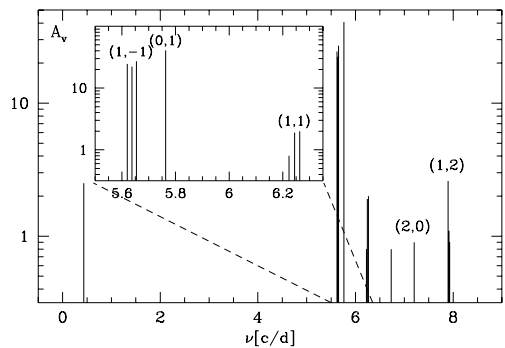}}
\caption{\footnotesize Oscillation spectrum of $\nu$ Eridani with
mode identification. Data are from \citet{jerz}. The main new
feature in comparison with data of \citet{hand} is the prograde
component of the (1,2) triplet. The mode identifications, shown as
the $(\ell,n)$ values with $n=-1$ denoting the g$_1$-mode, are
from \citet{deri}.}\label{nu}
\end{figure}

Here I will concentrate on one of these objects, $\nu$ Eridani,
the target of recent photometric \citep{hand,jerz} and
spectroscopic \citep{aerts} campaigns. Its oscillation spectrum is
reproduced in a schematic way in Fig. \ref{nu}. There is
identification of three modes, $\ell=0$, $p_1$, $\ell=0$, $p_1$
and $\ell=1$, $g_1$. For other two modes we know $\ell$ and $n$
but not $m$. Constraints on the star interior from the mode
frequencies were independently derived by \citet{pam} and by
\citet{aus}.

One significant seismic inference is a stringent limit on the
extent of convective overshooting ($\alpha_{\rm ov}\ge0.1$). The
limit relies mainly on frequency difference between the $\ell=1$,
g$_1$ and $p_1$ mode, which is very sensitive to the element
distribution in the layer outside the core. The difference between
p$_1$ $\ell=1$ and $\ell=0$ mode frequency is sensitive to the
metal abundance parameter, $Z$.

The other inference concerns the rotation rate in the same layer
outside the core and it is based on the rotational splitting of
the two $\ell=1$ modes. Though the frequency difference is small
(the models are near the avoided crossing), the difference in the
kernels linking the splitting to the local rate, $\Omega(r)$, is
large. Both kernels have maxima in the g- and p- mode propagation
zones but their relative size differs. The g$_1$ mode splitting is
much more sensitive to the rotation rate in the former zone, which
encompasses the chemically inhomogenuous zone outside the core. It
was found that the layer must rotate faster than envelope (factor
2.5 according to recent data).

The problem posed by the data is explaining mode excitation over
the unusually wide frequency range. The seismic model yields
$Z=0.015$ and with the standard solar heavy element mix this
implies mode instability in the frequency range from 4 to 6 c/d.
All dominant modes are in this range but there is a definite
detection of mode at 0.43  c/d and few modes above the upper
limit. An {\it ad hoc} enhancement of the iron abundance by factor
4 in the opacity bump layer solves the problem \citep{pam} but
plausibility  of such an enhancement has yet to be checked by
means of evolutionary model calculations.

\section{Final remarks}

There has been some progress in exploitation of rich oscillation
spectra in the context of unsolved problems of stellar interior
physics. New constraints on element diffusion and mixing were
derived from data on sdB and $\beta$ Cep stars. Constraints on
convection in outer layers were obtained from seismic data on a
$\delta$ Scuti-type star. Available frequency data on solar-like
pulsators are not yet accurate enough for similar applications but
there are very interesting prospects for probing deep interior in
moderate mass stars with expected more accurate data on $\eta$ Boo
and Procyon.

Mode frequencies are not only seismic observables of interest. For
unstable mode pulsators, important are simultaneous multicolor
photometric and spectroscopic data on individual modes  which are
essential for constraining mode identification and, in addition,
provide additional constraints on stellar modeling. For solar-like
pulsators, data on peak amplitudes and widths provide unique
information on subphotospheric convection. As \citet{sam} showed
current models do not explain data for $\alpha$ Cen.

Understanding oscillation spectra means not only mode identification but
also explaining origin and diversity in the form of observed oscillations.
The very presence of oscillations sets a constraint on stellar
physics as best demonstrated in the case of sdB pulsators \citep{charpinet}.
New data on $\nu$ Eridani \citep{hand} cast some doubts
on current understanding of the linear driving mechanism in
$\beta$ Cep stars. Perhaps also in this case the same effect of iron accumulation
as proposed for sdB pulsation must be invoked.
Progress in nonlinear modeling is needed to explain
puzzling features of oscillation spectra found in some opacity driven
pulsators and, in particular, in the most multimode $\delta$ Scuti star FG
Vir.

\begin{acknowledgements}
I am grateful to Alosha Pamyatnykh for his help in preparing this
paper. The work was supported by MNiI grant No. 1 PO3D 021 28
grant
\end{acknowledgements}

\bibliographystyle{aa}

\end{document}